\documentclass[twocolumn,pra,10pt,showpacs]{revtex4}
\usepackage{graphicx,amsmath,amssymb}

\newcommand{\beq}{\begin{equation}}
\newcommand{\eeq}{\end{equation}}
\newcommand{\beqa}{\begin{eqnarray}}
\newcommand{\eeqa}{\end{eqnarray}}
\newcommand{\ket}[1]{| #1 \rangle}
\newcommand{\bra}[1]{\langle #1 |}

\newcommand{\opda}{\hat{\Delta}_{\textrm{A}}(\varphi_{\textrm{A}})}
\newcommand{\opdb}{\hat{\Delta}_{\textrm{B}}(\varphi_{\textrm{B}})}
\newcommand{\opdab}{\hat{\Delta}(\varphi_{\textrm{A}}, \varphi_{\textrm{B}})}
\newcommand{\opdg}{\hat{\Delta}_{\gamma}(\varphi)}
\newcommand{\opsp}{\hat{S}_+}
\newcommand{\opsm}{\hat{S}_-}
\newcommand{\opi}{\hat{I}}
\newcommand{\opio}{| 1 \rangle\langle 0 |}
\newcommand{\opoi}{| 0 \rangle\langle 1 |}
\newcommand{\twopiinverse}{\frac{1}{2 \pi}}
\newcommand{\density}{\hat{\rho}}
\newcommand{\opl}{\hat{\Lambda}(\varphi_+,\varphi_-)}
\newcommand{\ppm}{\mathcal{P}(\varphi_+,\varphi_-)}
\newcommand{\g}{\Gamma}
\newcommand{\gp}{\Gamma_+}
\newcommand{\gm}{\Gamma_-}
\newcommand{\gmax}{\Gamma_{{\rm \mbox{sup}}}}
\newcommand{\uab}{\hat{U}_{\textrm{AB}}}
\newcommand{\ua}{\hat{U}_{\textrm{A}}}
\newcommand{\ub}{\hat{U}_{\textrm{B}}}
\newcommand{\phiplus}{\Phi^+}
\newcommand{\phiminus}{\Phi^-}
\newcommand{\psiplus}{\Psi^+}
\newcommand{\psiminus}{\Psi^-}

\def\black{}

\def\sugg#1{$\lbrack$#1$\rbrack$\black}
\begin{document}


\title{A measurable entanglement criterion for two qubits}

\author{Hoshang Heydari}
\email{hoshang@imit.kth.se} \homepage{http://www.ele.kth.se/QEO/}
\affiliation{Department of Microelectronics and Information
Technology, Royal Institute of Technology (KTH), Electrum 229,
SE-164 40 Kista, Sweden}

\author{Gunnar Bj\"{o}rk}
\affiliation{Department of Microelectronics and Information
Technology, Royal Institute of Technology (KTH), Electrum 229,
SE-164 40 Kista, Sweden}

\author{Luis L. S\'{a}nchez-Soto}
\affiliation{Departamento de \'{O}ptica,
Facultad de Ciencias F\'{\i}sicas,
Universidad Complutense, 28040 Madrid, Spain}

\date{\today}

\begin{abstract}
We propose a directly measurable criterion for the entanglement of
two qubits. We compare the criterion with other criteria, and we
find that for pure states, and some mixed states, it coincides
with the state's concurrence. The measure can be obtained with a
Bell state analyser and the ability to make general local unitary
transformations. However, the procedure fails to measure the
entanglement of a general mixed two-qubit state.
\end{abstract}

\pacs{42.50.Hz, 42.50.Dv, 42.65.Ky}

\maketitle

\section{Introduction}

Entanglement, by which we mean the existence of nonlocal quantum
correlations between spacelike separated systems, is one of the
most fertile and thought-generating properties of quantum
mechanics. It has become an essential resource for the new fields
of quantum physics created in recent years
\cite{Pre98,Gru99,NiC00,Key02}, with some potential applications
such as quantum cryptography \cite{Bennett84,Ekert91} and quantum
teleportation \cite{Bennett93}. But the idea of quantum
entanglement is not new, it goes back to the early days of quantum
theory where it was initiated by Einstein, Podolsky and Rosen
\cite{EPR35} and was later extended by Bell \cite{Bell64} in form
of Bell inequalities.

During recent years, there has been ample activity to quantify the
existence and the degree of entanglement. The quest has been
directed to find measures that satisfy reasonable axioms, such as
normalization, continuity, monotonicity, etc. \cite{Don01}, and
that hopefully are simple, directly measurable, and valid for any
quantum state in any dimension. The foundation was laid by Peres,
who showed that a necessary criterion for separability was
positivity of the density matrix upon partial transposition
\cite{Peres96}. Soon thereafter, the Horodeckis proved that the
criterion was also sufficient \cite{Hor96}. In principle, the
criterion also showed a way to decide, experimentally, whether the
members of an identical ensemble of bipartite states were
separable or not. A direct test is out of the question, since
partial transposition is a nonphysical process. However, to make
an indirect test, one could first perform quantum tomography on
the ensemble to reconstruct its associated density matrix.
Subsequently, the positivity of the partial transpose of the
estimated density matrix could be checked. A problem of this
approach is that quantum state tomography is a resource-consuming
task. Although it has been performed on {\it a priori} known
states \cite{Raymer,Wineland,Lvovsky}, extending the method to
{\it a priori} unknown states, multipartite states, or highly excited states will be a difficult
task. Neverteless, recently there has been an experimental
demonstration of the method, where the polarization entanglement
of two light beams was determined by state tomography
\cite{Barbieri}. In view of the experimental difficulties with
(non-)separability tests via tomography, it is tempting to try to
use Bell tests, but such schemes have been shown to fail, in
general, since there exist classes of nonseparable states that do
not violate any Bell inequality \cite{Werner}.

The result of Peres and Horodeckis  spurred others to derive
quantitative measures of entanglement for nonseparable states,
such as entanglement of formation \cite{Bennett96}, distillable
entanglement \cite{Bennett962}, concurrence \cite{Wootters98}, or
relative entropy of entanglement \cite{Vedral97}. The problem with
these measures, again, is that they are difficult to measure.

Other methods to measure the entanglement of unknown states have
also been suggested. However for pure bipartite states it has been
shown that there exists no single observable that can measure
entanglement \cite{Sancho}. Optimal strategies involving local
measurements on individual members of an identically prepared
ensemble have been derived \cite{Sancho}, as well as optimal
measurements if one measures properties of the entire ensemble
(coherent measurements) \cite{Acin}. In both cases, it suffices to
make local measurements on the two systems, but again, the
measurements impose substantial experimental difficulties.

Entanglement witnesses is another method of detecting entanglement
\cite{Lewen00}. The witness operator is such that it gives a
negative expectation value for its associated (class of) entangled
state(s). The disadvantage with the witnesses is that a positive
expectation value does not assure separability. Recently,
entanglement witnesses that require only local measurements have
been proposed \cite{Guhne,Guhne2}. It is to be noted that
entanglement witnesses have been experimentally implemented very
recently \cite{Bourennane}.

Finally, a method that characterizes the entanglement for any
bipartite state has been proposed \cite{Horodecki}. This method is
based on what is referred to as a structural approximation of the
linear (antiunitary) transposition operator. Although, applicable
to any bipartite state, the method is based on a POVM that seems
difficult to implement experimentally.

In this paper, we develop a new measure for $2 \otimes 2$
bipartite quantum states, inspired by the measure of entanglement
proposed in \cite{Soto02}. Our measure equals the concurrence for
pure states and some classes of mixed states, and it is based on
the bipartite phase sum and difference as quantified by a positive
operator-valued measure (POVM) of quantum ``phase''. Although
tailor-made POVMs (e.g., POVMS optimal for some purpose) are
notoriously difficult to implement experimentally, it is shown
that our measure can be obtained from a Bell-state analyzer, a
single nonlocal observable. However, the measurement is
supplemented by variable local unitary transformations,
effectively rendering the measurement a multi-observable scheme.
From the experimental viewpoint, such a measurement is, by
comparison, relatively easy to implement.

\section{Joint phase properties of two qubits}

As stated in \cite{Soto02}, the most general POVM for a single
qubit
\begin{eqnarray}
\opdg & = & \twopiinverse (\opi + \gamma e^{i \varphi} \opsp +
\gamma e^{-i \varphi} \opsm ) \nonumber \\
& = & \twopiinverse (\opi + \gamma e^{i \varphi} \opio + \gamma
e^{-i \varphi} \opoi ) ,
\end{eqnarray}
where $0 \leq \gamma \leq 1$ to ensure positivity. It is easy to
see that for any value of $\gamma$ we have $\int_0^{2 \pi} d
\varphi {\rm Tr} [ \density_{\textrm{A}} \opdg ] = 1$, where
$\density_{ \textrm{A}}$ is an arbitrary qubit density operator,
so that ${\rm Tr} [ \density_{ \textrm{A}}\opdg ]$ is the
probability distribution function of the qubit ``phase''
$\varphi$. Only the choice $\gamma = 1$ enables the probability
distribution to attain the value zero, so in the following we will
assume that $\gamma =1$ and suppress this parameter in the
notation. Suppose that we have two qubits, denoted A and B. It is
now possible to form a POVM of this bipartite system by simply
forming the tensor product \beq \opdab = \opda \otimes \opdb ,
\eeq where,  e.g., $\varphi_{\textrm{A}}$ is the POVM ``phase''
associated with qubit A. We can now recast this POVM expressed in
local properties in terms of the phase difference $\varphi_-
\varphi_{\textrm{A}} - \varphi_{\textrm{B}}$ and sum $\varphi_+
\varphi_{\textrm{A}} + \varphi_{\textrm{B}}$, both of which are
joint properties of the qubits. Moreover, to make the associated
POVM $2 \pi$-periodic we define \cite{Luis}
\begin{eqnarray}
\opl & = & \frac{1}{2} \left [ \hat{\Delta}(\frac{\varphi_+ +
\varphi_-}{2},\frac{\varphi_+ - \varphi_-}{2}) \nonumber \right . \\
& & \left . + \hat{\Delta}(\frac{\varphi_+ +
\varphi_-}{2}+\pi,\frac{\varphi_+ - \varphi_-}{2}+\pi)\right ] .
\end{eqnarray}
From this joint POVM, we can naturally form the associated joint
probability distribution \beq \ppm =  \mathrm{Tr} \mbox{[}
\density  \opl  \mbox{]} , \eeq where $\density$ is the two-qubit
density operator. Next, we introduce measures of to what extent
the properties of the bipartite system depend on the joint phase
properties. To that \sugg{end} we compute \beq \gp = \left |
\int_{2 \pi} d\varphi_+ e^{i \varphi_+} \ppm \right | \eeq and
\beq \gm = \left | \int_{2 \pi} d\varphi_- e^{i \varphi_-} \ppm
\right | \mbox{,} \eeq which are the Fourier components of the
joint phase probability distribution. Note that, due to the small
size of the two qubit Hilbert space, $\ppm$ does not contain any
Fourier components of higher frequency. Lastly, we define \beq \g
= 4 \pi |\gp - \gm| . \eeq Computing this value explicitly for a
two-qubit state, with its density operator $\density$ expressed in
the standard qubit basis $\ket{00}$, $\ket{01}$, $\ket{10}$,
$\ket{11}$, one obtains \beq \g = 2 \left ||\rho_{14}| -
|\rho_{23}| \right| . \eeq This is our central result. From this
equation one can immediately deduce that $0 \leq \g \leq 1$. It is
also clear that for any separable state, $\density \density_{\textrm{A}} \otimes
\density_{\textrm{B}}$, one gets
$\g=0$ since $\rho_{14}=\rho_{A12}\rho_{B12}$ and
$\rho_{23}=\rho_{A12}\rho_{B12}^\ast$. However, the measure $\g$
is not invariant under local unitary transformations (and the
measure $\mathbb{D}$ in \cite{Soto02} suffers from the same
deficiency). This can easily be seen by looking at the Bell state
$\ket{\phiplus} = (\ket{00} + \ket{11})/\sqrt{2}$. Its density
matrix, expressed in the standard qubit basis, is \beq \density \frac{1}{2}\left (
\begin{array}{cccc}
1 & 0 & 0 & 1 \\
0 & 0 & 0 & 0 \\
0 & 0 & 0 & 0 \\
1 & 0 & 0 & 1
\end{array}
\right ) \eeq from which one directly gets $\g = 1$, indicating
two maximally entangled qubits. However, by making the local
unitary transformation $\ket{0} \rightarrow (\ket{0} +
\ket{1})/\sqrt{2}$, $\ket{1} \rightarrow (\ket{0} -
\ket{1})/\sqrt{2}$ to qubit A only, the density matrix becomes
\beq \density = \frac{1}{4}\left (
\begin{array}{cccc}
1 & 1 & 1 & -1 \\
1 & 1 & 1 & -1 \\
1 & 1 & 1 & -1 \\
-1 & -1 & -1 & 1
\end{array}
\right ). \eeq For this density matrix we get $\g = 0$ (and
similarly, the measure $\mathbb{D}$ in \cite{Soto02} is zero too,
for this matrix). The reason for this result can loosely be stated
as follows: We have based our measure on joint phase properties.
However, there exists a complementary property to phase, which is
excitation. (More precisely, relative phase, and relative
excitation are complementary operators\cite{Luis93,Bjork03}). The
joint properties with respect to this complementary quantity are
not probed by $\g$. Hence, to get the complete information of the
joint properties of the state, we need to make local
transformations such that the entanglement, which is invariant
under local unitary transformations, is manifested in the phase
properties. \textit{Hence, we define our measure of entanglement}
$\gmax $, \textit{where} sup \textit{refers to the supremum of}
$\g$ \textit{under all possible local unitary transformations.} It
follows directly from the definition that this measure is
invariant of what local bases the density matrix is expressed in.

\section{An experimental implementation}

In the remainder of this paper we shall compare it to some of the
existing entanglement measures and we shall demonstrate how the
measure can be experimentally measured. First, let us look at a
pure two qubit state $\ket{\psi} = \alpha_1 \ket{00} + \alpha_2
\ket{01} + \alpha_3 \ket{10} + \alpha_4 \ket{11} $. For pure
states, the concurrence, which is an established measure of
entanglement, is $C = 2 |\alpha_1 \alpha_4 - \alpha_2 \alpha_3|$
\cite{Wootters98}. (The separability criterion $\alpha_1 \alpha_4
- \alpha_2 \alpha_3=0$ has also been derived from a Hilbert space
geometrical viewpoint in \cite{Moss01}.) Our measure $\g$, on the
other hand, is $\g=2 ||\alpha_1 \alpha_4^\ast| - |\alpha_2
\alpha_3^\ast|| \leq C$. Equality is only achieved if either the
complex vectors $\alpha_1 \alpha_4$ and $\alpha_2 \alpha_3$ are
parallel, or if at least one of the coefficients $\alpha_1, \ldots
, \alpha_4$ equals zero. Assume this is not the case. We then want
to find $\gmax$. To this end we perform local unitary
transformations $\ua \otimes \ub = \uab$ on the qubits A and B.
Assume that the transformation yields new state coefficients
$\alpha_1', \ldots , \alpha_4'$. The local transformations leave
the concurrence \textit{C} invariant, so the relation $\g \leq C$
still holds. If it is always possible to fulfill this relation
with equality, then $\gmax = C$ for pure states. This is indeed
always true. Assume,  e.g., that we want to find a $\uab$ that
makes $\alpha_1'=0$ (making $\g = C$). The most general local
unitary transformation will yield
\begin{eqnarray} \alpha_1'
 & = & \alpha_1 \cos \varphi \cos \vartheta + \alpha_2 e^{i \theta_{\textrm{B}}}
\cos \varphi \sin
 \vartheta \nonumber \\
& &  + \alpha_3 e^{i \theta_{\textrm{A}}}\sin \varphi \cos
\vartheta + \alpha_4 e^{i(\theta_{\textrm{A}} +
\theta_{\textrm{B}})} \sin \varphi \sin
 \vartheta,
\end{eqnarray}
where $\theta_{\textrm{A}}$ and $\theta_{\textrm{B}}$ are relative
phases between the states in qubit A and B, respectively, and the
angles $\varphi$ and $\vartheta$ determine the mixing between the
two states of each qubit. Choosing $\varphi = 0$, $\vartheta
\arctan|\alpha_1/\alpha_2|$, and $\theta_{\textrm{B}} = \arg
(\alpha_1/\alpha_2)+ \pi$ will ensure that $\alpha_1'=0$ so that
$\g = \gmax = C$. Hence, for all pure states our criterion gives
identical results than the concurrence.

 To find an experimental scheme to obtain our measure, we note that if we
find some local unitary transformations that maximizes $\g$, it is
possible to simultaneously make $\rho_{14}$ and $\rho_{23}$ real
by locally changing the relative phase between the qubit states.
All such phase changes will leave $|\rho_{14}|$ and $|\rho_{23}|$
invariant. That is a transformation of the type \beq \left (
\begin{array}{cc}
1 & 0 \\
0 & e^{i \theta_{\textrm{A}}}
\end{array}
\right ) \otimes \left (
\begin{array}{cc}
1 & 0 \\
0 & e^{i \theta_{\textrm{B}}}
\end{array}
\right ) , \eeq can make $\rho_{14}$ and $\rho_{23}$ real and
positive simultaneously. When this is the case, we note that $2
\rho_{14} = \bra{\phiplus}\density \ket{\phiplus} -
\bra{\phiminus}\density \ket{\phiminus}$, and $2 \rho_{23}
\bra{\psiplus}\density \ket{\psiplus} - \bra{\psiminus}\density
\ket{\psiminus}$. From these relations, the strategy becomes
clear. We measure the state in a Bell-state analyzer and look at
the difference between the detection probabilities $P_{\phiplus} +
P_{\psiplus} - P_{\phiminus} - P_{\psiminus}$. The maximum of this
difference under all possible local unitary transformations of the
state gives $\gmax$. (We could, of course instead measure,  e.g.,
$P_{\phiplus} + P_{\psiminus} - P_{\phiminus} - P_{\psiplus}$, as
this function also reaches the maximum $\gmax$ when $\rho_{14}$ is
real and positive and $\rho_{23}$ is real and negative. Hence,
$\gmax$ is essentially the optimal entanglement visibility
\cite{Jaeger,Volz}. Needless to say, such a measurement of
entanglement would not be simple to perform on an unknown state,
since we need to find the maximum over four continuous (but
bounded) parameters. However, simulations we have made show that
in most cases it suffices to maximize over each of the four
parameters in turn, to find the global maximum. This is due to the
fact that $\g$ is a nicely behaved function for many classes of
states. Making the local unitary transformations is relatively
easy for,  e.g., polarization-entangled photon pairs. In this
case, the necessary transformations can be accomplished with
geometric rotations and birefringent optics.

\section{Mixed states}

For mixed states the situation is more complicated and, in
general, our measure will fail. We have already noted above that
if the density matrix is a direct tensor product of any two qubit
density matrices, then $\gmax=0$, as it should. The standard test
of entanglement measures for mixed states is to look at the
mixture of Bell, or Bell-like states. One such mixture, used by
the Horodeckis \cite{Hor96}, is the state \beq \density = p
\ket{\psi_1} \bra{\psi_1} + (1-p) \ket{\psi_2} \bra{\psi_2} , \eeq
where $0 \leq p \leq 1$, $\ket{\psi_1} = a \ket{00} + \sqrt{1-a^2}
\ket{11} $ and $\ket{\psi_2} = a \ket{01} + \sqrt{1-a^2} \ket{10}
$. It is relatively straightforward to show that $\gmax = 2 a
(1-a^2)^{1/2}|1 - 2 p |$. That is, the state is nonseparable for
$a \neq 0,1$ or $p \neq 1/2$. Another state that is often used for
comparison is the Werner state
\begin{eqnarray} \density & = & p \ket{\Psi_-} \bra{\Psi_-} +
(1-p) (\ket{\Psi_+} \bra{\Psi_+} \nonumber \\ & & +
\ket{\phiminus} \bra{\phiminus} + \ket{\phiplus} \bra{\phiplus})/3
. \label{Werner type}\end{eqnarray} Again, it is not hard to
compute that $\gmax = |1 - 2 p|$. That is, the state is separable
for $p=1/2$, and maximally entangled for $p = 1$. In both these
cases our measure agrees with the concurrence. In fact, any state
that can be written as a diagonal matrix in the Bell-basis
$\ket{\phiplus},\ket{\phiminus},\ket{\psiplus},\ket{\psiminus}$
(with eigenvalues $\lambda_1, \ldots , \lambda_4$) will have the
form \beq \density = \frac{1}{2} \left (
\begin{array}{cccc}
\lambda_1+\lambda_2 & 0 & 0 & \lambda_1-\lambda_2 \\
0 & \lambda_3+\lambda_4 & \lambda_3-\lambda_4 & 0 \\
0 & \lambda_3-\lambda_4 & \lambda_3+\lambda_4 & 0 \\
\lambda_1-\lambda_2 & 0 & 0 & \lambda_1+\lambda_2
\end{array}
\right ), \eeq when written in the standard basis. From this we
can easily compute that if two of the eigenvalues are zero, our
measure $\gmax$ equals the concurrence $C$. If three or more
eigenvalues are non-zero, our method fails, in general. Our method
also fails for general mixed states of the form $\density = \sum_n
P_n \ket{\psi_n}\bra{\psi_n}$. Hence, unless one has some
\textit{a priori} information of the state,  e.g., that it is of
the class expressed in (\ref{Werner type}) or that it is pure, our
proposed method is not reliable.

\section{Discussion and conclusion}

In conclusion we have proposed a way of \textit{measuring} the
concurrence of any two-qubit pure state. An advantage with the
method is that it effectively only requires one observable,
although the state must be subjected to local unitary
transformations before the measurement to determine its
entanglement. A disadvantage compared to other proposed schemes is
that the needed observable is nonlocal, meaning that one has to
have access to both qubits. The method also works for some mixed
states commonly used to test measures of entanglement, but fails
for general mixed states. This serves as a warning that one has to
be cautious when testing entanglement measures or measurement
procedures with these states.

\begin{acknowledgments}
This work was supported by the Swedish Research Council (VR) and
the Swedish Foundation for Strategic Research (SSF). The authors
acknowledge useful discussions with Professor A. Karlsson and for
helpful comments by Dr. P. Marsden.
\end{acknowledgments}

\end{document}